\documentclass[12pt]{article}
\begin{document}
\vskip 0.1cm

\centerline{\large \bf SPIN EFFECTS IN TWO QUARK SYSTEM}
\centerline{\large \bf AND MIXED STATES }

\vskip 0.7cm \centerline{I.Haysak$^{\ast}$, Yu.Fekete$^{\ast}$,
V. Morokhovych $^{\ast}$, S.Chalupka $^{\ast \ast} $,M. Salak
$^{\ast \ast \ast}$}

\vskip .2cm \centerline{\sl Department of Theoretical Physics,
Uzhgorod National University,}
 \vskip .2cm \centerline{\sl 32
Voloshyna Str., Ukraine, 88000}

\vskip .2cm \centerline{\sl Department of Theoretical Physics and
Geophysics, Safarik University,}

 \vskip .2cm \centerline{\sl 16
Moyzesova Str., 041 54 Kosice, Slovak Republic}
 \vskip .2cm \centerline{\sl Department of
Physics, Presov University,}
 \vskip .2cm \centerline{\sl  17 Novembra Str., 080 09 Presov,
Slovak Republic}

\vskip 1.0cm

\begin{abstract}
Based on the numeric solution of a system of coupled channels for
vector mesons  ($S$-  and $D$-waves mixing) and for tensor mesons
($P$- and $F$-waves mixing)  mass spectrum and wave functions of
a family of vector mesons $q\overline{q}$  in triplet states are
obtained. The calculations are performed using a well known
Cornell potential with a mixed Lorentz-structure of the
confinement term. The spin-dependent part of the potential is
taken from the Breit-Fermi approach. The effect of singular terms
of potential is considered in the framework of the perturbation
theory and by a configuration interaction approach (CIA),
modified for a system of coupled equations. It is shown that even
a small contribution of the $D$-wave  to be very important at the
calculation of certain characteristics of the meson states.

\end{abstract}

\vskip .3cm

\vskip 8cm

\hrule

\vskip .5cm

\noindent \vfill $ \begin{array}{ll} ^{\ast}\mbox{{\it e-mail :}}
& \mbox{haysak@univ.uzhgorod.uzhgorod.ua}
\end{array}
$

$ \begin{array}{ll} ^{\ast}\mbox{{\it e-mail :}} &
 \mbox{feketeyu@univ.uzhgorod.ua}
\end{array}
$
$ \begin{array}{ll} ^{\ast}\mbox{{\it e-mail :}} &
 \mbox{morv@univ.uzhgorod.ua}
\end{array}$

$ \begin{array}{ll} ^{\ast}\mbox{{\it e-mail :}} &
 \mbox{chalupka@pjsu.sk}
\end{array}$

\vfill \baselineskip=13pt

\section{Introduction}
Meson states, which are considered as the bound states of a
quark-antiquark system, are convenient objects for studying both
the strong interaction effects and various characteristics of weak
interaction [1,2]. For the description of the low-energy
properties of hadrons the following approaches are used:
Bete-Solpiter approach, lattice calculation technique and
potential models. Each of these methods has its own advantages and
shortcomings. The potential models are the simplest from the
point of view of mathematics, which is essential for practical
calculations. In the framework of the present model the averaged
mass spectrum [1], spin effects [3,4] and the decay widths of
heavy quarkonia [5] can be well described. As concerning the
light-quark systems, the situation is rather contradictory. Using
the same parameters, some of the effects (spin effects, decay,
averaged mass spectrum) can be described [6], but not all the
effects together. The reason for this is the fact the light-quark
systems being explicitly relativistic, and relativistic potential
models should be applied for them.

When the spin effects are considered in potential models (in
Breit-Fermi approach), singular terms of the interquark
interaction potential of the form $\frac{1}{r^{3}}$  and $\delta(
{\bf r})$ arise. This is a serious problem at the calculations of
the meson characteristics. As a rule, in such case perturbation
theory is used [7].  But it has certain shortcomings. The main
disadvantage is that the theory supposes small interaction, but
in hadron physics it is the perturbation of the order 30-500 MeV,
what is comparable with the distance between the unperturbed
energy levels, hence, the condition of small perturbation not
being fulfilled.The variational method also very popular in hadron
physics [8]. These methods were in fact the first, which used to
investigate the characteristics of mesons, considered as
two-quark systems.

In the triplet state configurations with uncertain orbital moment
$L=J+1$, $L=J-1$ occur. For example, mixing of $S$- and $D$-waves
occurs in the state $1^{-}$ and $P$- and $F$- mixing for the state
$2^{+}$ (we use the spectroscopic notations $J^{P}$  where the
total moment and the system parity are indicated). In most papers
where the triplet states are considered (See Ref. 3 and references
therein) the authors neglect the channel coupling or introduce an
additional parameter - the mixing angle [9-11].

    In papers [12] the mixed state $1^{-}$ for $q\overline{q}$-systems was
considered in the frame of coupled channel and mass spectra with
lepton decay width was investigated. It was observed a small
$D$-wave mixing, but this small contribution causes essential
influence on the value of hyperfine splitting and decay width.

 In the present paper the effect of the singular terms
of the interquark interaction potential for the quarkonium state
$1^{-}$, described by a system of coupled equations, is studied
and the comparison of perturbation methods and configuration
interaction, modified for the coupled equation system, is
performed. We followed Ref. [12], when choose the Lorentz
structure and parameters of $q\overline{q}$-interaction with the
numerical solution scheme of the coupled differential equation.
The calculation version of the configuration interaction approach
based on the Schroedinger equation is taken from Ref. 3, where the
hyperfine splitting is considered. In such approach the mixing
angle for $S$-  and $D$-waves is determined by the system dynamics
and does not require any additional experimental data.

\section{Breit-Fermi approach}
    It is widely accepted that the interaction between two quarks
(or heavy quark-antiquark) consist of a short range part
describing the one-gluon exchange and a infinitely rising
long-range part responsible for confinement of the quarks

\begin{equation}
\label{eq1}V_{0}(r)=-\frac{\alpha_{S}}{r}+\beta r.
\end{equation}

Wilson loop techniques suggest that the confining potential
should be purely scalar, but relativistic potential calculations
which have been published [13-15] show a need for (some) vector
confinement. Maintaining $V_{V}(r)+V_{S}(r)$ unchanged we do
allow a fraction of vector confinement [6,13,14]. Namely

\begin{equation}
\label{eq2}V_{V}(r)=-\frac{\alpha_{S}}{r}+\beta_{V}r, \ \ \
V_{S}(r)=\beta_{S}r \ \ \    (\beta_{V}+\beta_{S}=\beta )
\end{equation}

The confining potential transforms as the Lorentz scalar and
vector potential transforms as the time component of a
four-vector potential. As we can see, the choice of Lorentz
structure of potential for quark-antiquark interaction is
important model for study of spin effects [7,14-16].

We consider the Breit-Fermi Hamiltonian for case $m_{1}=m_{2}=m$

\begin{equation}
\label{eq3}H=\frac{p^{2}}{m}+V_{0}+H_{LS}+H_{SS}+H_{T}
\end{equation}

with spin-orbit term

\begin{equation}
\label{eq4}H_{LS}=\frac{1}{2mr^{2}}\left(3\frac{dV_{V}}{dr}-\frac{dV_{S}}{dr}\right)\bf{(LS)}
\end{equation}

where $\bf{S}=\bf{S}_{1}+\bf{S}_{2}$ is the total spin of bound
state and $\bf{L}$ is the relative angular momentum of its
constituents, the spin-spin term

\begin{equation}
\label{eq5}H_{SS}=\frac{2}{3m^{2}}\Delta V_{V}(r)\bf{S_{1}S_{2}}
\end{equation}

and the tensor term

\begin{equation}
\label{eq6}H_{T}=\frac{1}{12m^{2}}\left(\frac{1}{r}\frac{dV_{V}}{dr}-\frac{d^{2}V_{S}}{dr^{2}}\right)S_{12}
\end{equation}

where

\begin{equation}
\label{eq7} S_{12}=12\left(\frac{{\bf
(S_{1}r)(S_{2}r)}}{r^{2}}-\frac{1}{3}{\bf S_{1}S_{2}}\right)
\end{equation}

For bound state constituents of spin
${\bf{S}_{1}}={\bf{S}_{2}}=\frac{1}{2}$, the scalar product of
their spin, $\bf{S_{1}S_{2}}$, is given by ${\bf
S_{1}S_{2}}=-\frac{3}{4}$ for singlet states ($S=0$), and ${\bf
S_{1}S_{2}}=\frac{1}{4}$ for triplet states ($S=1$).

Taking into account (2), then (4), (5) and (6) yield

\[
V_{LS} = \frac{1}{2 m_{q} m_{\overline{q}}} \left( 3
\frac{\alpha_{S}}{r^{3}} + 3 \frac{\beta_{V}}{r} -
\frac{\beta_{S}}{r} \right) (\bf{LS})
\]

\[
V_{SS}=\frac{4}{3m_{q}m_{\overline{q}}}\left(\frac{\beta_{V}}{r}-2\pi\alpha_{S}\delta(r)\right)(\bf{S_{1}S_{2}})
\]

\[
V_{T}=\frac{1}{m_{q}m_{\overline{q}}}\left(\frac{\alpha_{S}}{r^{3}}+\frac{\beta_{V}}{r}\right)S_{12}
\]

The physical states of $q\bar q$ system are determined by total
momentum $J$ and its projection $M_{J}$, parity $P$ and total spin
$S$ (Table 1), here we use spectroscopic notation $^{2S+1}L_{J}$

 Table 1.States of two fermion system

\begin{center}
\begin{tabular}{|c|c|c|c|c|}
\hline & \multicolumn{2}{c|}{Singlet states} &
\multicolumn{2}{c|}{Triplet states} \\
& \multicolumn{2}{c|}{$(S=0)$} &
\multicolumn{2}{c|}{$(S=1)$} \\
\hline
$J/P$ & + & - & +  & - \\
\hline
0   &  - - - -  & $^1S_0$  & $ ^3P_0$  & - - - -  \\
\hline
1 & $^1P_1$ & - - - -     & $^3P_1$ & $^3S_1 + ^3D_1$ \\
\hline
2 & - - - - & $^1D_2$   & $^3P_2 + ^3F_2$ & $^3D_2$ \\
\hline
\end{tabular}
\end{center}

It is seen from Table 1 that there are singlet and triplet states
with definite orbital moment and triplet states with mixed
orbital components ($L=J\pm1$). For pure states wave functions are
of the form $\left(F(r)/r\right){\cal Y}_{JLS}^{M}$ where ${\cal
Y}_{JLS}^{M}$ is the spin-orbital part of the wave function.

The radial part of wave function $F(r)$ for singlet states
satisfies equation

\begin{equation}
\label{eq8}
\frac{d^{2}F}{dr^{2}}+[k^{2}-\frac{L(L+1)}{r^{2}}-^{1}\upsilon_{c}]F=0
\end{equation}

where $^{1}\upsilon_{c}=m^{1}V_{0}; \ \
^{1}V_{0}=V_{V}(r)+V_{S}(r)-(3/4)V_{SS}$ and $k^{2}=mE$.

The radial function of the pure triplet states $^{3}P_{0}$ and
states with $L=J$ obeys the same equation that singlet states
(8), but with potentials
$^{3}V_{0}=V_{V}+V_{S}+\frac{1}{4}V_{SS}-2V_{LS}-4V_{T}$ and
$^{3}V_{0}=V_{V}+V_{S}+\frac{1}{4}V_{SS}-V_{LS}+2V_{T}$
respectively.

The wave function for ground triplet state of $q\overline{q}$
system with negative parity ($P=-1$) is a mixture of states
$^{3}S_{1}$ and $^{3}D_{1}$ and may be put in the form

\begin{equation}
\label{eq9} \psi=\psi_{S}+\psi_{D}\equiv\frac{1}{r}u(r){\cal Y
}_{101}^{1}+\frac{1}{r}w(r){\cal Y}_{121}^{1}
\end{equation}

Then the equation $(H - E) \psi = 0$ is equivalent to coupled
system [17]

\[
[ - \frac{1}{m} \frac{d^2}{dr^2} - E + ^3V_c] u + \sqrt 8 V_T w =
0,
\]

\begin{equation}
[ - \frac{1}{m} \frac{d^2}{dr^2} - E + \frac{6}{m r^{2}} + ^3 V_c
- 2 V_T - 3 V_{LS}] w + \sqrt 8 V_T u = 0, \label{eq10}
\end{equation}

where

\[
^3V_c = V_V + V_S + \frac{1}{4} V_{SS}.
\]

The Schroedinger equations are linked due to the presence of a
tensor component $V_{T}$  in the interaction potential. In Ref. 12
the system (10) was solved numerically for regular part of the
potential (4-6), and irregular terms were calculated as a first
order of the perturbation theory. Hereinafter we study the
contribution of the singular terms in the framework of the
perturbation theory and CIA. We generalize these methods to the
system of equations. For this purpose that is convenient to
rewrite the system (10) in a matrix form. The Hamiltonian of the
system can be separated to regular and irregular parts $ H=H_{0}+W
$, namely

\begin{equation}
\label{eq11}\widehat{H_{0}}=\left(\begin{array}{cc}
  -\frac{1}{m}\frac{d^{2}}{dr^{2}}-\frac{\alpha}{r}+kr+\frac{\beta_{V}}{3m^{2}r}; & \sqrt{8}\frac{\beta_{V}}{m^{2}r}; \\
   \sqrt{8}\frac{\beta_{V}}{m^{2}r}; &
    -\frac{1}{m}\frac{d^{2}}{dr^{2}}-\frac{\alpha}{r}+kr+\frac{\beta_{V}}
   {3m^{2}r}+\frac{6}{mr^{2}}-\frac{2\beta_{V}}{m^{2}r}-\frac{3}{m^{2}}(\frac{3\beta_{V}-\beta_{S}}{r});
\end{array}\right)
\end{equation}

and

\begin{equation}
\label{eq12}\widehat{W}=\left(\begin{array}{cc}
  \frac{2\pi\alpha}{3m^{2}}\delta(\overrightarrow{r}); & \sqrt{8}\frac{3\alpha}{m^{2}r^{3}}; \\
  \sqrt{8}\frac{3\alpha}{m^{2}r^{3}}; & \sqrt{8}\frac{3\alpha}{m^{2}r^{3}}
  -\frac{6\alpha}{m^{2}r^{3}}-\frac{9\alpha}{2m^{2}r^{3}};
\end{array}\right)
\end{equation}

Then unperturbated eigenfunction of equation

 \begin{equation}
 \label{13}H_{0}\Psi_{n}=E_{n}\Psi_{n}
 \end{equation}
 where

 \begin{equation}
\label{eq14}\Psi_{n}=\left(\begin{array}{c}
  u_{n} \\
  w_{n}
\end{array}\right)
\end{equation}
can be used as a basis for the full wave function of the system,
namely,

\begin{equation}\label{eq15}
\Phi=\sum a_{n}\Psi_{n}
\end{equation}

Then the matrix elements of the perturbated term

\begin{equation}
\label{eq16}
W_{mn}=\int\Psi_{m}^{*}\widehat{W}\Psi_{n}d\overrightarrow{r }
\end{equation}

are given by

\begin{equation}
\label{eq17} W_{mn}=\int
u_{m}(-\frac{2\pi\alpha}{3m^{2}}\delta(\overrightarrow{r}))u_{n}dr+
\int w_{m}(\sqrt{8}\frac{3\alpha}{m^{2}r^{3}})u_{n}dr+ \int
u_{m}(\sqrt{8}\frac{3\alpha}{m^{2}r^{3}})w_{n}dr+ \\ \\
\end{equation}
\[
 +\int w_{m}(-\frac{2\pi\alpha}{3m^{2}}\delta(\overrightarrow{r})-
\frac{6\alpha}{m^{2}r^{3}}-\frac{39\alpha}{2m^{2}r^{3}})w_{n}=
W_{SS}^{u_{m}u_{n}}+ W_{ST}^{w_{m}u_{n}}+ W_{ST}^{u_{m}w_{n}}+
W_{SS+LS+ST}^{w_{m}w_{n}}
\]

and in the framework of the first order of perturbation theory the
correction to the energy spectrum is equal to

\begin{equation}
\label{eq18}\Delta E_{nn}=W_{SS}^{u_{n}u_{n}}+2
W_{ST}^{w_{n}u_{n}}+W_{SS+LS+ST}^{w_{n}w_{n}}
\end{equation}

It can be noted that mean value of unperturbed Hamiltonian
$H_{0}$ (11) consist of the same parts determined by $S$-wave,
$SD$-interferention and $D$-wave correspondent
\[
E^{(0)}=\langle\psi|H_{0}|\psi\rangle=E^{S}+E^{SD}+E^{D}.\ \ \ \
\ \ \  \ \ \ \ \ \ \ \ \ \ \ \ \ \ \ \ \ \ \ \ \ \ \ (18a)
\]
 As seen from Eq.
(17), the $S$-wave gives pure contribution only for the spin-spin
interaction (the first term), the interferential $SD$ term (the
second and third ones) contains the spin-tensor correction, and
pure $D$-wave (the fourth term) includes all the spin-dependent
components of the interaction potential. This structure is
reflected in Eq. (18) by sub- and superscripts. In the case of
the configuration interaction method we used an algorithm,
developed in Ref. 3,4. The mean value $E$ of Hamiltonian
$H_{0}+W$ is an eigenvalue of the following system of linear
algebraic equations:

\begin{equation}\label{eq19}
\begin{array}{c}
   a_{1}(E-E_{1}^{0}-W_{11})-a_{2}W_{12}-a_{3}W_{13}-...-a_{n}W_{1n}=0 \\
  -a_{1}W_{21}+a_{2}(E-E_{2}^{0}-W_{22})-a_{3}W_{23}-...-a_{n}W_{2n}=0 \\
  .................................................................... \\
  -a_{1}W_{n1}-a_{2}W_{n2}-a_{3}W_{n3}-...+a_{n}(E-E_{n}^{0}-W_{nn})=0
  \\
\end{array}
\end{equation}

where $E^{0}_{i}$- are eigenvalues of Hamiltonian $H_{0}$ (11)
and $W_{ij}$ - are correspondent matrix elements (16).
Respectively, eigenvectors $(a_{1},a_{2},...,a_{n})$ gives
 eigenfunctions $\Phi_{i}$, $i=1,2,...,n$ of the
 Hamiltonian $H_{0}+W$.

\section{Hyperfine and fine splitting}

\hspace*{0.5cm}  To study the role of tensor forces we calculated
for triplet states $L=J\pm1$ in mixed form (10), and for single
$S$-wave (as most authors do). It was pointed [14,16] that the
best agreement with experimental data for potential (1) is
obtained when $\beta = 0.18 GeV^{2}$. For description of
$q\overline{q}$-system we were varying parameter $\beta_V $ (and
fixing $\beta_V + \beta_S = 0.18 GeV^2)$ ) to achieve the
agreement with experimental mass splitting of $1^{--}$ states.
Finally we have used the following parameters: $\beta_V = 0.001
GeV^2, \beta_S = 0.179 GeV^2$ for $u \bar u$-systems; $\beta_V =
0.04 GeV^2, \beta_S = 0.14 GeV^2$ for charmonium and bottonium,
$\alpha_s (b \bar b) = 0.24, \alpha_s (c \bar c) = 0.38, \alpha_s
(u \bar u) = 0.54$. The quark masses are: $m_b = 4.7 GeV$ $m_c =
1.4 GeV$, and $m_u = 0.33 GeV$. Tables 2 - 4 list numerical
results, namely, mass spectra for single $S$-wave and mixed
state, fraction of tensor part of the potential to energy levels
(18a) and $D$-wave fraction in the total wave function
($P_{D}=\int|w|^{2}dr$). And for comparison we showed experimental
values [19] and
calculation with screened potential[3,4]. \\

 {\bf Table 2.}{\bf  Hyperfine splitting for the
charmonium}
\begin{center}
\begin{tabular}{|c|c|c|c|c|c|c|}
\hline State & $S$-wave & $SD$-waves & [3] & [19] & $E^{SD},$
&$P_D$,\\
& $E_{theor}, MeV$ & $E_{theor}, MeV$ & $E_{theor},
MeV$ & $E_{exp}, MeV$ & \% & \% \\
\hline
$1^1 S_0$ & 2980 & & & 2980 & & \\
\hline
$1^3 S_1$ & 3153 & 3097 & & 3097 & 16 & 0.05  \\
\hline
$1^3 S_1 - 1^1 S_0$ & 173 & 117 & 110 & 117 & & \\
\hline
$2^1 S_0$ & 3642 &  &  & 3590 & & \\
\hline
$2^3 S_1$ & 3759 & 3734 &  & 3685 & 3 & 0.8 \\
\hline
$2^3 S_1 - 2^1 S_0$ & 117 & 92 & 67 & 95 & & \\
\hline
$3^1 S_0$ & 4107 & & & - & & \\
\hline
$3^3 S_1$ & 4208 & 4192 & & 4040 & 1 & 1.3 \\
\hline
$3^3 S_1 - 3^1 S_0$ & 101 & 85 &  & -  &  & \\
\hline
\end{tabular}
\end{center}

{\bf Table 3.}{\bf  Hyperfine splitting for the bottonium}
\begin{center}
\begin{tabular}{|c|c|c|c|c|c|c|}
\hline State & $S$-wave & $SD$-waves & [3] & [19] & $E^{SD},$
&$P_D$,\\
& $E_{theor}, MeV$ & $E_{theor}, MeV$ & $E_{theor},
MeV$ & $E_{exp}, MeV$ & \% & \% \\
\hline
$1^1 S_0$ & 9415 & & & - & & \\
\hline
$1^3 S_1$ & 9462 & 9460 & & 9460 & 2.1 & 0.004  \\
\hline
$1^3 S_1 - 1^1 S_0$ & 47 & 45 & 46 & - & & \\
\hline
$2^1 S_0$ & 9883 &  &  & - & & \\
\hline
$2^3 S_1$ & 9911 & 9911 &  & 10023 & 0.2 & 0.04 \\
\hline
$2^3 S_1 - 2^1 S_0$ & 28 & 28 & 26 & - & & \\
\hline
$3^1 S_0$ & 10201 & & & - & & \\
\hline
$3^3 S_1$ & 10224 & 10224 & & 10355 & 0.1 & 0.1 \\
\hline
$3^3 S_1 - 3^1 S_0$ & 23 & 23 &  & -  &  & \\
\hline
\end{tabular}
\end{center}

 {\bf Table 4.}{\bf  Hyperfine splitting for $(u \vec
u)$-systems}

\begin{center}
\begin{tabular}{|c|c|c|c|c|c|c|}
\hline State & $S$-wave & $SD$-waves & [3] & [19] & $E^{SD},$
&$P_D$,\\
& $E_{theor}, MeV$ & $E_{theor}, MeV$ & $E_{theor},
MeV$ & $E_{exp}, MeV$ & \% & \% \\
\hline
$1^1 S_0$ & 140 & & & 140 & & \\
\hline
$1^3 S_1$ & 674 & 640 & & 770 & 4 & 0.02  \\
\hline
$1^3 S_1 - 1^1 S_0$ & 534 & 500 & 923 & 630 & & \\
\hline
$2^1 S_0$ & 1134 &  &  & 1300 & & \\
\hline
$2^3 S_1$ & 1564 & 1543 &  & 1450 & 1 & 0.04 \\
\hline
$2^3 S_1 - 2^1 S_0$ & 430 & 409 & 411 & 150 & & \\
\hline
\end{tabular}
\end{center}

\hspace*{0.5cm} It can be noted that we obtain
 good description of mass spectra and hyperfine splitting for heavy quark systems.
 For light system our quasirelativistic model is less success, here relativistic
 kinematics play sufficient role [6].\\
\hspace*{0.5cm} In Tables 5-7 fine splitting data are listed for
$P$-states, where we indicate in brackets result with mixed tensor
states ($^{3}P_{2}-^{3}F_{2}$ mixing). Comparison results for
$S-D$ and $P-F$ system show that influence of tensor forces
decreasing when total moment of the system increases.

\par
{\bf Table 5.}{\bf Fine splitting for $c\overline{c}-$ systems (in
MeV)}

\begin{tabular}{|c|c|c|c|c|c|c|c|}
\hline \textbf{State} & $\triangle M$ & Our results & [7] &
[20] &[21] & [22] & [19] \\
\hline $1 P$ & $M(^{3}P_{2})-M(^{3}P_{1})$ & 51(49) & 576 & 51 &
45
& 28 &$45.67$  \\
& $M(^{3}P_{1})-M(^{3}P_{0})$ & 72 & 76 & 83 & 91 & 32 &
$95.51$ \\
& $M(^{3}P_{2})-M(^{3}P_{0})$ & 123(121) & 132 & 134 & 137 & 66 &
 $141.18$\\
 \hline
\end{tabular}

\par

{\bf Table 6.} {\bf Fine splitting for $b\overline{b}-$ systems
(in MeV)}

\begin{tabular}{|c|c|c|c|c|c|c|c|}
\hline \textbf{State} & $\triangle M$ & Our results &[7] &
[20] & [21] & [21] & [19] \\
\hline $1 P$ & $M(^{3}P_{2})-M(^{3}P_{1})$ & 12(11) & 23 & 31 & 24
& 28 & $19.9 \pm 1.1$  \\
& $M(^{3}P_{1})-M(^{3}P_{0})$ & 15 & 26 & 41 & 37 & 32 &
$32.8 \pm 1.5$ \\
& $M(^{3}P_{2})-M(^{3}P_{0})$ & 27(26) & 49 & 72 & 61 & 66 & $52.7
\pm 1.5$ \\
\hline $2 P$ & $M(^{3}P_{2})-M(^{3}P_{1})$ & 10 (10) & 16 & 24 &
17
& 20 & $13.3 \pm 0.9$  \\
& $M(^{3}P_{1})-M(^{3}P_{0})$ & 15 & 20 & 32 & 26 & 24 &
$23.1 \pm 1.1$ \\
& $M(^{3}P_{2})-M(^{3}P_{0})$ & 25(25) & 36 & 56 & 44 & 44 &
$36.4\pm 1.0$ \\
\hline
\end{tabular}

\par

{\bf Table 7. Hyperfine splitting in $P-$waves (in MeV)}

\begin{tabular}{|c|c|c|c|c|c|c|}
\hline \textbf{State} & $\triangle M$ & Our results & [7] & [19] \\
\hline $c\overline{c}(1P)$ & $M(^{3}P_{1})-M(^{1}P_{1})$ &$-10$ & 13 & $-15.63 \pm 0.36$  \\
$b\overline{b} (1P)$ & $M(^{3}P_{1})-M(^{1}P_{1})$ & -3.4 & 4.3 & $---$ \\
$b\overline{b} (2P)$& $M(^{3}P_{1})-M(^{1}P_{1})$ & -3.5 &  3 &
$---$ \\ \hline
\end{tabular}

\section{Vector mesons in CI approach}
We also analyzed influence of perturbation part of the system
Hamiltonian because most authors study spin-spin interaction in
the frame of perturbation theory of the first order. To estimate
high order theory we use the configuration interaction approach
(CIA), the one configuration case of which coincides with the
first order perturbation theory.

Below the numerical results for the mass spectrum of
$c\overline{c}$ and $u\overline{u}$ systems are given. The
results of the mass spectrum calculations listed in Table 8
(nonperturbated eigenvalues of energy, correction according to
 the perturbation theory, energies according to CIA with two and
 three configurations). It is seen from the table that the series
 of  Eq. (10) converges rapidly for heavy systems and  already the
  perturbation theory gives more than 90\% contribution of
  the singular terms, but for light mesons the extension beyond
  the perturbation theory is essential.

{ \bf Table 8. Vector meson mass spectrum}

  \begin{tabular}{|c|c|c|c|c|c|}
    \hline
    \textbf{States} &{\scriptsize  $\mathbf{M}_{TH}, MeV$ }&{\scriptsize  $\mathbf{M}_{TH},
    MeV$}&{\scriptsize  $\mathbf{M}_{TH}, MeV$} &{\scriptsize  $\mathbf{M}_{TH}, MeV$} & {\scriptsize $\mathbf{M}_{EXP}, MeV$} \\
   &{\scriptsize  \textbf{non-perturb.}} &{\scriptsize  \textbf{1 config.}} &{\scriptsize  \textbf{2 config.}} &{ \scriptsize \textbf{3 config.}} & \\
    \hline $J/\psi(1S)$ & $3043.45$ & $3098.58$ & $3097.11$ & $3096.87$ & $3096.87$  \\
    \hline $\psi(2S)$ & $3649.04$ & $3674.02$ & $3675.50$ & $3674.87$ & $3685.96$ \\
\hline $\psi(4040)$ & $4098.04$ & $4113.11$ & $---$ &  $4113.97$ & $4040$ \\
    \hline $\rho(770)$ & $685.5$ & $778$ & $771$ & $768.5$ & $768.5$ \\
\hline    $\rho(1450)$ & $1574.5$ & $1643$ & $1650$ & $1643$ & $1465$ \\
\hline    $\rho(1700)$ & $2270.5$ & $2332$ & $---$ & $2341$ & $1700$ \\
\hline
  \end{tabular}

{ \bf Table 9.Perturbation theory for a $J/\psi$ meson }

\begin{tabular}{|c|c|c|c|}
\hline    & $\mathbf{W_{SS}^{u_{n}u_{n}}, MeV}$ &
 $\mathbf{2W_{ST}^{w_{n}u_{n}}, MeV}$ & $\mathbf{W_{SS+ST+LS}^{w_{n}w_{n}}, MeV}$ \\
\hline $\mathbf{\Delta E_{11}, MeV}$ & $6.04$ & $49.6$ & $0.00050$ \\
\hline $\mathbf{\Delta E_{22}, MeV}$ & $4.18$ & $21.16$ & $0.00036$ \\
\hline $\mathbf{\Delta E_{33}, MeV}$ & $3.65$ & $11.95$ & $0.00052$ \\
\hline
\end{tabular} \\

\par
{ \bf Table 10.Perturbation theory for a $\rho$ meson}

\begin{tabular}{|c|c|c|c|}
\hline    & $\mathbf{W_{SS}^{u_{n}u_{n}}, MeV}$ &
 $\mathbf{2W_{ST}^{w_{n}u_{n}}, MeV}$ & $\mathbf{W_{SS+ST+LS}^{w_{n}w_{n}}, MeV}$ \\
\hline $\mathbf{\Delta E_{11}, MeV}$ & $60.98$ & $31.67$ & $-0.19$ \\
\hline $\mathbf{\Delta E_{22}, MeV}$ & $50.19$ & $19.02$ & $-0.28$ \\
\hline $\mathbf{\Delta E_{33}, MeV}$ & $46.64$ & $14.69
$ & $-0.075$ \\
\hline
\end{tabular}

\noindent
  The absolute values of corrections due to certain singular terms of the potential in the
  framework of the perturbation theory are rather interesting. The correction values for
 the spin-spin, spin-tensor and spin-orbit components of the singular part of the interaction
  for charmonium and $\rho$ meson are given in Tables 9 and 10, respectively. It is seen that for
  heavy systems the spin-tensor part of the correction value by order of magnitude exceeds
 the spin-spin part and is totally determined by the presence of the $D$ wave admixture. For
 light systems the spin-spin and spin-tensor correction values are of the same order. Hence,
 when the spin effects are considered, one should take into account the orbital structure of
 the meson states, especially for the systems of heavy quarks. However, we note that the $D$-wave
 admixture for the light and heavy systems is less than 1 \% and about 4\% ,
 respectively [12]. In comparison with Ref. 10
 , we note that for the $\Psi(2S+D)$  state the $D$-wave admixture
 in our case is $P_{D}=0.008$  [16], which corresponds to the mixing angle of $\varphi= 5^{o}$. In Ref. 7
  the mixing angle of pure triplet states $2S$  and $1D$  $\varphi=12^{o}$  is quoted.

\section{\bf Leptonic decay of heavy quarkonia and wave
function in the origin}

\hspace*{0.5cm} For the leptonic decay widths of two-quark system
we shall consider decay of vector states into $e^+ e^-$ pairs.
The leptonic decay width of system $M_{q \bar q}\to e^+ e^-$ is
calculated from the  Van Royen-Weisskopf  formula [23]

\begin{equation}
\tilde \Gamma (^3S_1 \to e^+ e^-) = \frac{4 \alpha^2_{em}Q^2}
{M_{q \bar q}^{2}}|R(0) |^2, \label{eq20}
\end{equation}

where $M_{q \bar q}$ is mass of vector meson, $Q$ is quark charge,
$\alpha_{em}$ is the fine structure constant and $R(0)$ is the
radial wave function in the origin. In our case radial wave
function in the origin is determined by $S$-wave component of the
system wave function, namely, by $\frac{u(r)}{r}$
(9) because the $D$-wave component vanishes in the origin.\\
\hspace*{0.5cm} The formula (20) is based on the notion that
constituent quark-antiquark pair annihilates into a single
virtual photon, which subsequently gives rise to a leptonic pair.
The relativistic and radiative QCD corrections [24] modifies
eq.(20)

\begin{equation}
\Gamma (^3 S_1 \to e^+ e^-) = \tilde \Gamma (1 - \frac{16
\alpha_s (m_{q}^2 )}{3 \pi}). \label{eq21}
\end{equation}

\hspace*{0.5cm} As Eichten and Quigg have pointed out [25] the
$QCD$ correction reduces the magnitude of $\Gamma$ significantly,
however the value of reduction is somewhat uncertain . For vector
mesons containing light quarks this formula leads to paradoxes
[26]. In paper [27] Motyka and Zalewski extrapolated eq. (21) by
rational and exponential function and obtain averaged formula.

\begin{equation}
\Gamma_{V \to e^+ e^-} = F(q) \frac{32 \alpha_{s}}{9 M^2_V}
|R(0)|^2, \label{eq22}
\end{equation}

with $F(c) = 4.73 . 10^{-5}$ for charmonium and $F(b) = 2.33 .
10^{-5}$ for bottonium. We have calculated decay widths using the
formula of Van Royen-Weisskopf (20) and formula (22). Table 11
lists this results.

{\bf Table 11. The leptonic decay widths of heavy mesons}
\begin{center}
\begin{tabular}{|c|c|c|c|c|c|c|}
\hline
State & $S$-waves & $SD$-wave  & [27] & [5] & [27]& [19] \\
& $\Gamma_{theor.}$,keV &  $\Gamma_{theor.}$,keV &
$\Gamma_{theor.}$,keV & $\Gamma_{theor.}$,keV &
$\Gamma_{theor.}$,keV &  $\Gamma_{exp.}$,keV
\\
\hline $J/ \psi 1 S$ & 8.2 (5.63) &7.8 (5.41) & 4.5 & 4.24 & 8.0
& $5.26 \pm 0.37$  \\
$\psi'2 S$ & 4.0 (2.79) & 3.7 (2.59) & 1.9 & 1.81 & 3.7 &
$2.12 \pm 0.18$ \\
$\psi'' 3 S$ & 2.9 (2.01) & 2.6(1.82)& - - - & 1.22
& - - - & $0.75 \pm 0.15$ \\
\hline $\Upsilon 1 S$ & 1.2 (1.01) &1.14(0.96) & 1.36 & 0.85 &
1.7 & $1.32 \pm 0.04$ \\
$\Upsilon' 2 S$ & 0.63 (0.53) & 0.58(0.49) & 0.59 & 0.38 &
0.8 & $0.52 \pm 0.03$ \\
$\Upsilon'' 3 S$ & 0.49 (0.42) & 0.44(0.37) & 0.40 & 0.27 &
0.6 & $0.48 \pm 0.08$ \\
\hline
\end{tabular}
\end{center}
%************************************************************
The calculations of widths were performed with tensor forces and
without ones. Value of the widths which were calculated by formula
(\ref{eq22}) are given in parentheses.

{\bf Table 13. Wave functions in the origin $|R(0)|^{2}$.} (In
$GeV^{3}$)
\begin{center}
\begin{tabular}{|c|c|c|c|c|c|c|c|}
  \hline State &$|R(0)|^{2},$  &
$|R(0)|^{2},  $ & $|R(0)|^{2}, $ & $|R(0)|^{2}, $ & $|R(0)|^{2},$
   & $ |R(0)|^{2},$ & $  |R(0)|^{2},$ \\
   & non-perturb. & 1 config. & 2 config. &3
   config.
   & $[5]$ &  $ [25]$ &  $ [18]$\\
  \hline
  $J/\psi(1S)$ & $0.77$ & $0.69$ & $0.83$ & $.85$ & $1.05$ & $1.45$ & $0.81$ \\
  \hline $\psi(2S)$ & $0.53$ & $0.56$ & $0.47$ & $0.50$ & $0.63$ & $0.93$ & $0.53$ \\
  \hline $\psi(4040)$  & $0.47$ & $0.53$ & $---$ & $0.41$ & $0.52$ & $0.79$ & $0.46$ \\
  \hline $\rho(770)$ & $0.11$ & $0.086$ & $0.126$ & $0.136$ & $---$ & $---$ & $---$ \\
  \hline $\rho(1450)$ & $0.091$ & $0.081$ & $0.073$ & $0.089$ & $---$ & $---$ & $---$ \\
  \hline $\rho(1700)$ & $0.084$ & $0.12$ & $---$ & $0.061$ & $---$ & $---$ & $---$ \\ \hline
\end{tabular}
\end{center}
The values of the squared radial wave function in the origin for
the non-perturbed and the perturbed case along with the CIA
calculations with two and three configuration sets are listed in
Table 13. It is seen that, contrary to the energy spectra, the
consideration of the decay widths beyond the framework of the
perturbation theory results in the correction value of 20-30\% .
Our results are very close to those of Ref. 17 where the same
quark systems were considered on the base of Schroedinger
equation with the generalized Breit-Fermi potential. The
difference from our work is in the choice of the Lorentz
structure of the quark-quark interaction and neglecting the
$D$-wave admixture.
    Also squared mean value radius obtained from our wave function
coincide well with other calculations (Table 12).

 {\bf Table 12. Squared mean
value radius}
\begin{center}
\begin{tabular}{|c|c|c|c|c|}
\hline
$nL$ & [22] & Our results  & [29] & Our results \\
& $\langle r^2_c \rangle^{1/2},$ fm & $\langle r^2_c
\rangle^{1/2}, $fm & $\langle r^2_b \rangle^{1/2},$ fm & $\langle
r^2_b
\rangle^{1/2},$ fm \\
\hline
1 SD & 0.43 & 0.433 & 0.24 & 0.256 \\
\hline
2 SD & 0.85 & 0.847 & 0.51 & 0.552 \\
\hline
3 SD & 1.18 & 1.182 & 0.73 & 0.768 \\
\hline
\end{tabular}
\end{center}

\hspace*{0.5cm} It must be noted that squared mean value radius
0.7 fm meet the requirements of quark-antiquark pair creation $(u
\bar u)$ (string break). It is necessary to modify the potential
model taking into account the opening of a new channel for those
conditions where the given value is overlapped.\\

\section{Conclusions}

In the present work we have studied the influence of the
quark-antiquark interaction potential structure to the meson
properties. We pay the main attention  to the role of the tensor
forces. It was considered mass spectra and leptonic decay width.

The regular part of the potential was taken into account by
numerical solutions of Schroedinger equations for singlet and
unmixed triplet states  but a coupled differential equation for
mixed triplet states ($S-D$- for vector mesons and $P-F$- for
tensor mesons). The irregular part of potential was taken into
account in the frame of the configuration interaction approach.

 The analysis of our results presented in Table 2
- 4 shows that differences between the theoretical calculations
of mass spectrum for heavy quarkonia and the experimental results
are 1 - 4 \%. Therefore we expect relativistic correction in the
range 1 - 16 \% for charmonium, and up to 1 \% for bottonium
spectrum. We have also been able to describe hyperfine splitting
of $c \bar c$- and  $b \bar b$-quarkonium. For describing mass
spectrum of light mesons it is necessary to use relativistic
potential model. We have calculated
hyperfine splitting $u \bar u$ system, too.  \\
It shown that, admixture of $D$-wave is small (less then 1\%) but
its influence to the mass spectra reach 16\%. The leptonic decay
widths suggest that for charmonium theoretical widths, which were
calculated by Van Royen-Weisskopf formula are systematically
higher than experimental data. But for bottonium we have
obtained  lower values. For $J / \psi$-meson better decay widths
are obtained by means of formula (\ref{eq21}), which takes into
account $QCD$ correction. Here the influence of $D$-waves ranges
from 4 \% for ground state up to 25 \% for the second excited
state, but for values calculated by Van Royen-Weisskopf formula
it is from 8 \% up to 50 \%. For $\Upsilon$-meson it is opposite.
More exact is the calculation done by (\ref{eq21}), but $QCD$
correction reduced the results more significantly. Besides,
$D$-waves contribute less than for charmonium: from 4 \% for
ground state up to  11 \% for
second excited state. \\
\hspace*{0.5cm} The results obtained show that contribution of
the $D$-waves is impossible to neglect for considered leptonic
decay width of quarkonia.

It is shown that, in spite of a small admixture of the  $D$-wave
this component of the wave function plays an essential role at
the account of the irregular part of the interaction potential,
and namely the irregular terms are considered by most authors
while the spin effects being studied. The presence of the $D$
-wave essentially enhances the contribution of the spin-tensor
component into the mass spectrum. As concerning the technique of
the account of the irregular part of the interaction it should be
noted that in case the mass spectrum of the systems being
considered, one can restrict themselves  by the first order
correction of the perturbation theory, but at the analysis of the
decay widths the CIA results essentially improve the first order
perturbation theory.

{\bf Acknowledgements}

 We would like to thank our colleagues for useful
discussions Lazur V.,  Haysak M.,  Shpenik A. and Rubish V.

\end{document}